\documentclass[a4paper,11pt]{article}

\usepackage{pos}
\usepackage{lipsum} %package to generate placeholder text in the following

\usepackage{tabto}
\usepackage{wrapfig}
\usepackage{graphicx}
\usepackage{amsmath}
\usepackage[noabbrev,capitalize]{cleveref}
\usepackage[tableposition=top]{caption}
\usepackage{subcaption}
\usepackage{adjustbox}
\usepackage{float}
\floatstyle{plaintop}
\restylefloat{table}

\title{Performance studies on new 4" photomultiplier types intended for IceCube-Gen2 optical modules}

\ShortTitle{IceCube-Gen2 PMT performance}

\author{The IceCube-Gen2 Collaboration \\{\normalsize 
\normalfont(a complete list of authors can be found at the end of the proceedings)}\\}

\emailAdd{markus.dittmer@uni-muenster.de}
\emailAdd{alexander.kappes@uni-muenster.de}

\abstract{

In the upcoming IceCube-Gen2 extension, the newly developed optical modules will include \linebreak[3]4--inch PMTs. For this purpose, the manufacturers Hamamatsu and North Night Vision Technology have developed new PMT models to meet the requirements of the IceCube-Gen2 science case. The specifications include strict requirements on temporal resolution, detection efficiency, and dark noise. We summarize the efforts to measure these performance characteristics and show that both PMT models meet the performance specifications set by IceCube-Gen2. Prototype optical modules based on both PMT models will be deployed with the IceCube Upgrade in order to test them in situ and help decide on a vendor for the Gen2 extension.
\vspace{4mm}
{\bfseries Corresponding authors:}
Markus Dittmer$^{1*}$, Alexander Kappes$^{1}$\\
{$^{1}$ \itshape University of Muenster}\\[4mm]
$^*$ Presenter

\ConferenceLogo{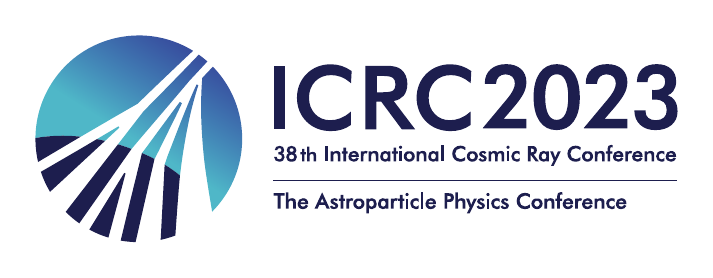}

\FullConference{The 38th International Cosmic Ray Conference (ICRC2023)\\ 26 July -- 3 August, 2023\\ Nagoya, Japan}
}

\begin{document}

\maketitle

\section{Introduction}\label{sec:intro}
\noindent
Photomultiplier-Tubes (PMTs) are the central component in Cherenkov telescopes such as BAIKAL-GVD \cite{BAIKAL}, KM3NeT \cite{KM3Net} and IceCube \cite{Aartsen2017}. In the planned extension of IceCube-Gen2 \cite{Aartsen2021}, 4" PMTs will be integrated into the optical modules to be embedded in the Antarctic ice. Specifically for this purposes, the proposed PMT models R16293-01-Y001 (hereafter referred to as BB) and N2041 (hereafter referred to as PO) were developed by Hamamatsu Photonics K.K (HPK) and North Night Vision Technology (NNVT), respectively. This study focuses on the third and final iteration of the PMT development process, excluding the results from earlier prototyping stages.
The performance of the utilized PMTs is a major contributor to the efficiency of the entire detector. Hence, strict requirements for various PMT characteristics have been set by the IceCube collaboration. The aim of the work described here is to verify that these specifications are met by conducting comparable measurements for both PMT models.
This work is structured into temperature dependent measurements (\cref{sec:temp}), the PMT quantum efficiency (\cref{sec:qe}) and homogeneity studies (\cref{sec:scan}).

\section{Temperature dependence}\label{sec:temp} %Performance with respect to temperature dependence

\begin{wrapfigure}{r}{0.45\textwidth}
\vspace{-20pt}
  \begin{center}
    \includegraphics[width=0.45\textwidth]{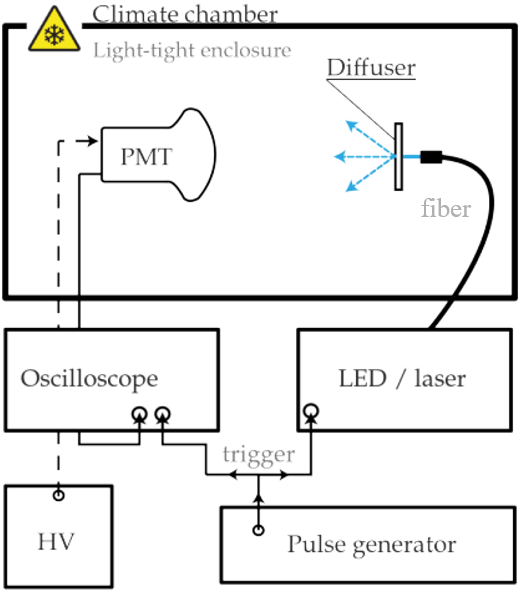}
  \end{center}
  \vspace{-20pt}
  \caption{Experimental setup for the measurement of PMT pulse parameters. Taken from \cite{MartinDoctorialThesis}.}
  \label{fig:TempSetup}
\end{wrapfigure}

\noindent
%Since the optical modules of IceCube are embedded in ice with a temperature ranging from -20\,°C to \linebreak -35\,°C \cite{Price2002}, it is crucial to examine the low temperature performance of the PMTs and its temperature dependency. 
%since the performance of a PMT can vary depending on the environmental temperature.
Since the optical modules of IceCube-Gen2 will be embedded in ice at different depths, with ambient temperatures ranging from $-8$\,°C to \linebreak $-40$\,°C \cite{Price2002}, it is crucial to examine the low temperature performance of the PMTs and its temperature dependency.

The measurements of this section were conducted using the setup depicted in \cref{fig:TempSetup}. It consists of a PMT %illuminated by a fibre equipped with a diffuser at 385\,nm \cite{Rongen2018}
illuminated by a LED at 385nm \cite{Rongen2018} through a fiber equipped with a diffuser
which is placed inside a light-tight enclosure. The enclosure is further housed inside a climate chamber which is ramped in 10\,°C steps from $-50$\,°C -- 20\,°C with the measurements being performed for several hours at each temperature. %The PMTs are driven by a passive base and operated with supply voltages of 700\,V -- 1000\,V, depending on the PMT gain. 
The PMT response is captured by an oscilloscope and the maximum amplitude, time of this amplitude and charge within a specified window are recorded for each waveform. For a more comprehensive understanding of the data acquisition, the reader is referred to \cite[pp.~50-52]{MartinDoctorialThesis}. With this setup, three different measurements were conducted for three Hamamatsu (BB9780, BB9786, BB9789) and three NNVT PMTs (PO4049, PO4052 PO4068) PMTs. % whereas two PMTs are measured simultaneously to save time.

\subsection{Gain calibration}
\begin{figure}[!t]
     \centering
     \begin{subfigure}[!htb]{0.45\textwidth}
         \centering
         \includegraphics[width=\textwidth]{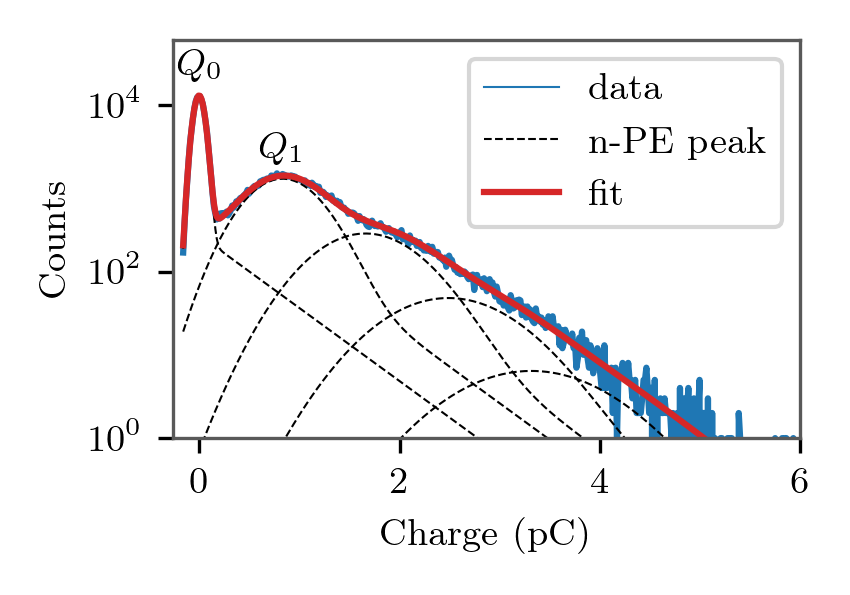}
     \end{subfigure}
     \hfill
     \begin{subfigure}[!htb]{0.45\textwidth}
         \centering
         \includegraphics[width=\textwidth]{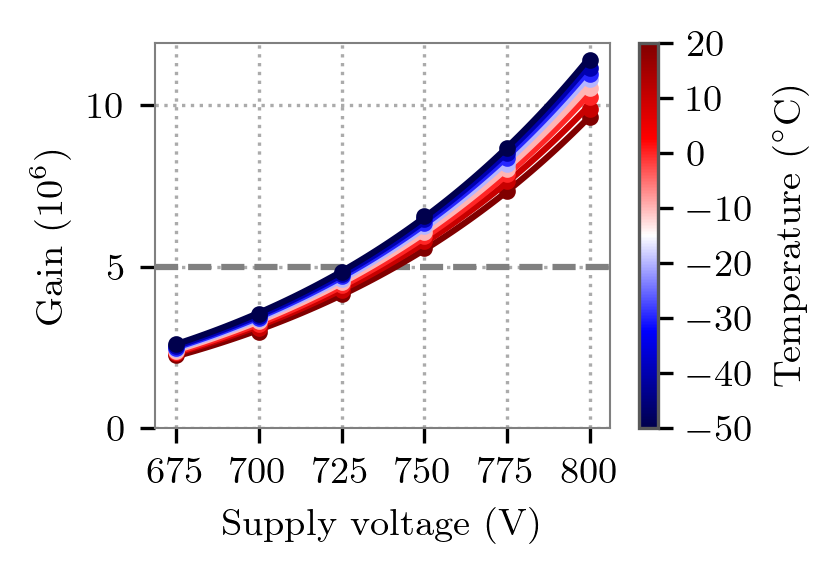}
     \end{subfigure}
        \caption{Example of SPE fit and gain curve for PO4049. The SPE fit of the charge histogram (left) provides the nominal gain, which represents one data point in the right plot. For each temperature, the nominal voltage (intersection with the gray line) is obtained.}
        \label{fig:Gain}
\end{figure}

\noindent
The first measurement with this setup aims at determining the supply voltage for which an amplification of $5 \cdot 10^6$ is achieved. This amplification level is referred to as \textit{nominal gain} and the corresponding supply voltage (dynode voltage ratio as recommended by the manufacturer) is known as \textit{nominal voltage}. %for which the dynode voltage ratio recommended by the manufacturer is applied.
%referred to as, pulses around the main response time of the PMT are measured % whereas the waveform sampling length is 150\,ns in respect to the LED pulse trigger. 
For this, the supply voltage is varied at each temperature. The charge spectrum measured for a given voltage is fit by the Single Photoelectron (SPE) formula as described \linebreak[3]in \cite[p.~45]{MartinDoctorialThesis}. An example of the SPE fit is shown in \cref{fig:Gain} (left).

\begin{wrapfigure}{r}{0.45\textwidth}
\vspace{-20pt}
  \begin{center}
    \includegraphics[width=0.45\textwidth]{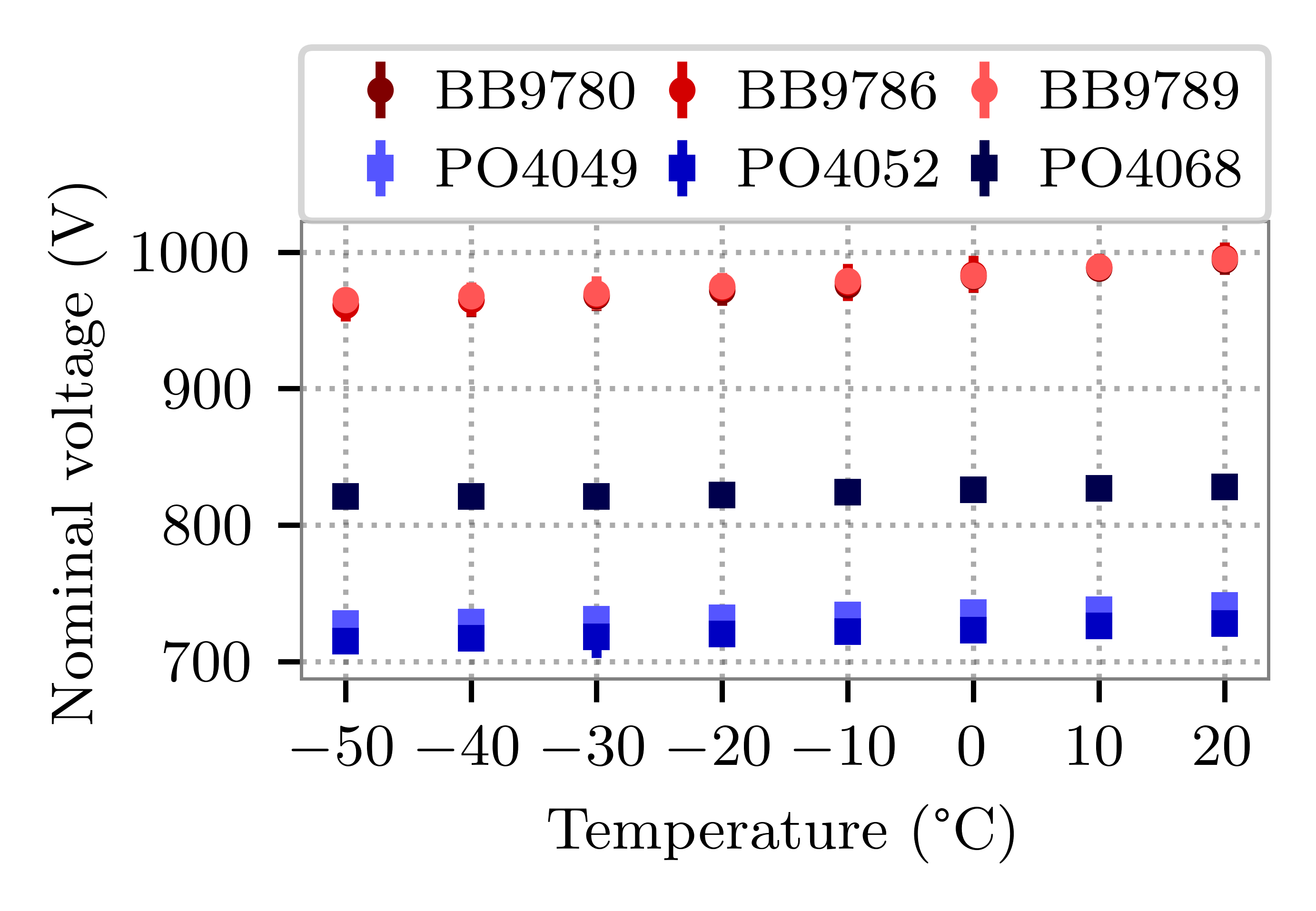}
  \end{center}
  \vspace{-20pt}
  \caption{Nominal voltages plotted against operation temperature. NNVT PMTs are plotted in red and Hamamatsu PMTs in blue colors.}
  \label{fig:NominalVoltage}
  \vspace{-10pt}
\end{wrapfigure}

\noindent
From this fit, the gain can be obtained from the charge of a single photoelectron ($Q_1/e$) amongst other parameters. The nominal voltage for each temperature of each PMT is obtained by fitting a power law of the form $c V^\beta$ \cite[p.~206]{Wright2017} which is shown in \cref{fig:Gain} (right) whereas the corresponding nominal voltages for all PMTs are presented in \cref{fig:NominalVoltage}.

Decreasing the temperature by $\Delta\text{T}=70\,$°C results in an average gain increase of $(29.1 \pm 0.4)\,\%$ for Hamamatsu and $(13.6 \pm 0.3)\,\%$ for NNVT PMTs. %reason for difference?
This temperature dependence of the gain is most likely attributed to an increase of the secondary emission coefficient of the dynodes. The obtained nominal voltages are set in the subsequent measurements.  
Furthermore, it is observed that the Hamamatsu PMTs exhibit a smaller spread in nominal voltage compared to the NNVT PMTs which on the other hand need a lower supply voltage to achieve the same gain. Although the sample size within this study is limited, manufacturer data of large production batches support this claim.% this claim is supported by data provided by the manufacturer. 

%(standard deviation 1.07\,V vs 45.72\,V) %This is however still expected since NNVT is still in the process of optimizing their manufacturing process as they started developing PMTs only quite recently.

%From this measurement, the rise time, fall time, pulse width, peak to valley ratio, SPE resolution, and light intensity can be obtained as well.

\subsection{Dark rate}
\begin{figure}[!htb]
     \centering
     \begin{subfigure}[!htb]{0.45\textwidth}
         \centering
         \includegraphics[width=\textwidth]{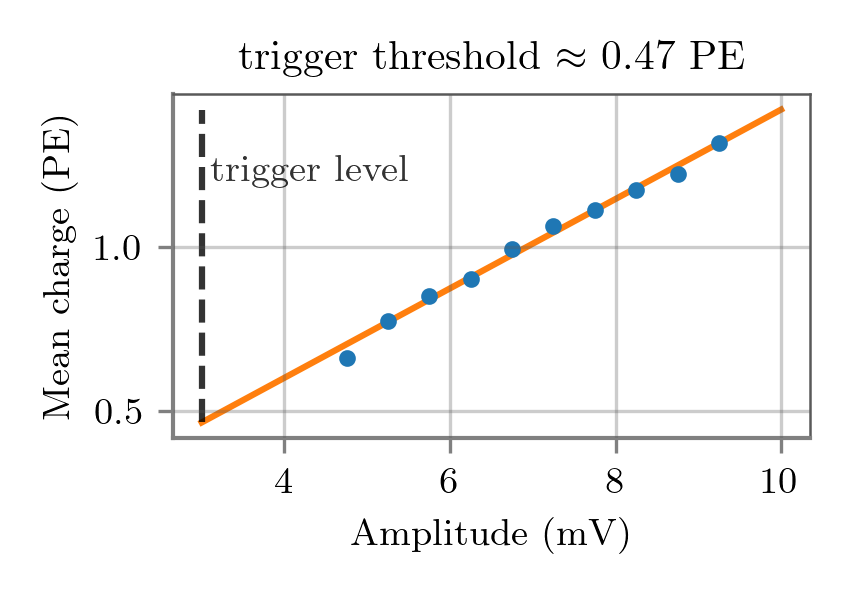}
     \end{subfigure}
     \hfill
     \begin{subfigure}[!htb]{0.45\textwidth}
         \centering
         \includegraphics[width=\textwidth]{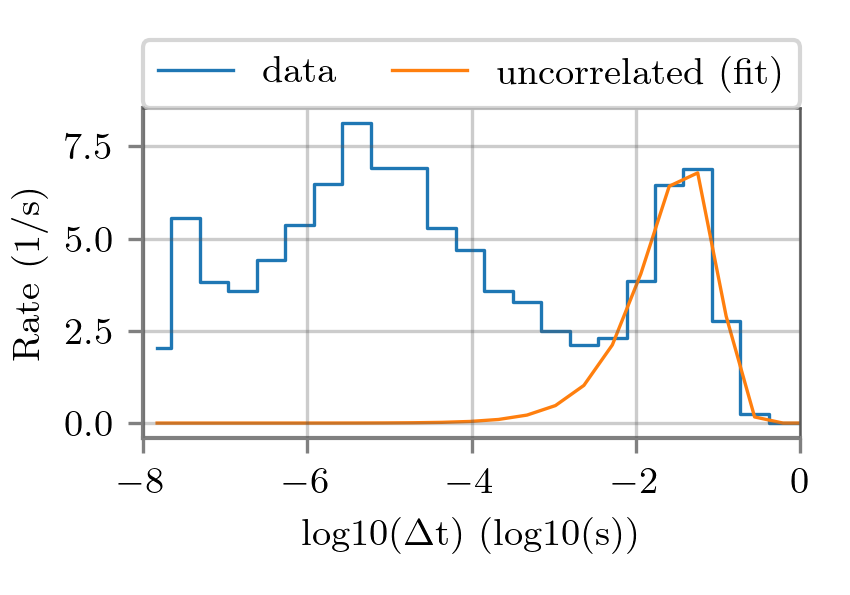}
     \end{subfigure}
        \caption{Additional evaluation for dark rates: (left) Binned amplitudes with the corresponding charge in units of PE where the trigger level in units of PE is obtained via a linear fit. (right) Fitting a Poissonian distribution onto the uncorrelated peak to classify noise contributions.}
        \label{fig:DR}
\end{figure}

\noindent
The dark rate refers to the intrinsic background noise characterized by counting pulse rates in the absence of any illumination. Since background pulses appear randomly in time, long waveforms of 1\,ms length were measured. From the number of pulses present in a given waveform one can calculate the dark rate which is done for more than $10^7$ waveforms for each temperature step. To avoid triggering on baseline noise, a threshold of 3\,mV ($\sim 0.47$\,PE) was set%meaning that this is the lowest value that a PMT pulse may be recorded with
. Consequently, low amplitude pulses are rejected. Since the charge is recorded and the nominal gain is known one can express this threshold in units of PE (charge$ / Q_1$) which is shown in \cref{fig:DR} (left). 

\noindent
In addition to this amplitude threshold, a charge cut of 0.2\,PE is set to further reject baseline noise. Note that a threshold--free measurement will produce higher dark rates than shown here. %However, the aim is the have a consistent comparison between all PMT within the specifications of the collaboration.

\begin{wrapfigure}{r}{0.45\textwidth}
\vspace{-20pt}
  \begin{center}
    \includegraphics[width=0.45\textwidth]{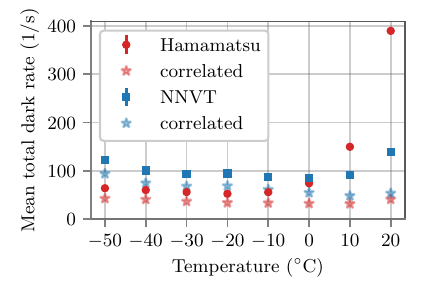}
  \end{center}
  \vspace{-20pt}
  \caption{Mean dark rates of both vendors plotted against temperature.}
  \label{fig:Darkrate}
  %\vspace{-20pt}
\end{wrapfigure}

A histogram based on the time difference between two successive pulses is shown in \cref{fig:DR} (right) where one can distinguish between correlated pulses caused by scintillation induced by radioactive decays in the glass bulb, and uncorrelated pulses, random pulses fitted by an exponential decay. For a more in--depth description, the reader is referred \linebreak[3]to \cite[pp.~108-112]{MartinDoctorialThesis}.

\cref{fig:Darkrate} displays the mean total dark rate of all PMTs from both vendors. Hamamatsu PMTs exhibit somewhat lower dark rates than NNVT PMTs at temperatures below 0\,°C. This difference is attributed to the reduced presence of correlated noise in Hamamatsu PMTs, indicating less scintillation induced by radioactive decays from the enveloping glass bulb. However, at temperatures above 0\,°C, the dark rates of Hamamatsu PMT increase significantly due to thermionic emission.
%Std from 3 PMTs below 0°C
%Hamamatsu: 4.478811848320826 NNVT: 20.20176726887706
%-30 and 20: 2.0687667437730703  15.727930657351164
%at -30 1.7988237455684983 13.519609664043763

\subsection{Timing resolution}
\noindent
To determine the timing characteristics, low light illumination (pulse occupancy less than 10\%; <5\,\% multi-PE pulses) is used such that mostly SPE pulses contribute -- otherwise the characteristics would yield different values in case multiple photons are detected simultaneously.  The timing resolution, known as Transit Time Spread (TTS), is calculated from 10-\textmu s-long waveforms.

\begin{figure}
     \centering
     \begin{subfigure}[!htb]{0.45\textwidth}
         \centering
         \includegraphics[width=\textwidth]{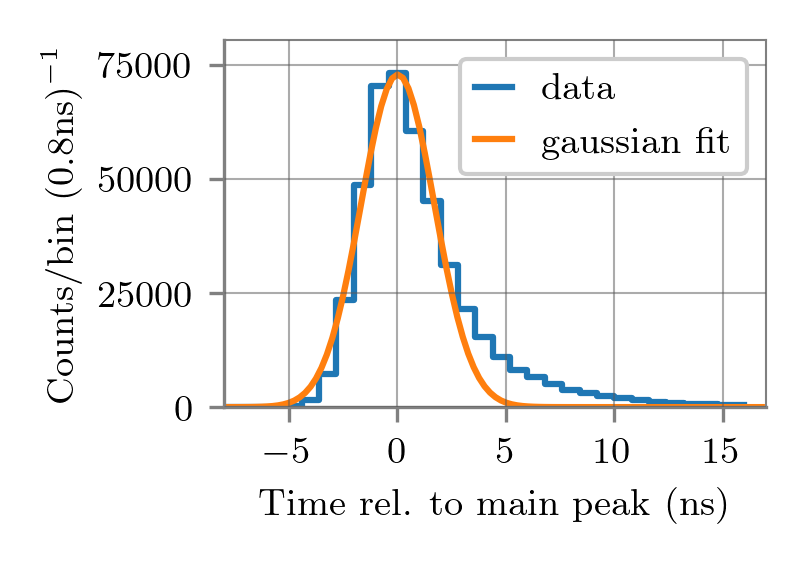}
     \end{subfigure}
     \hfill
     \begin{subfigure}[!htb]{0.45\textwidth}
         \centering
         \includegraphics[width=\textwidth]{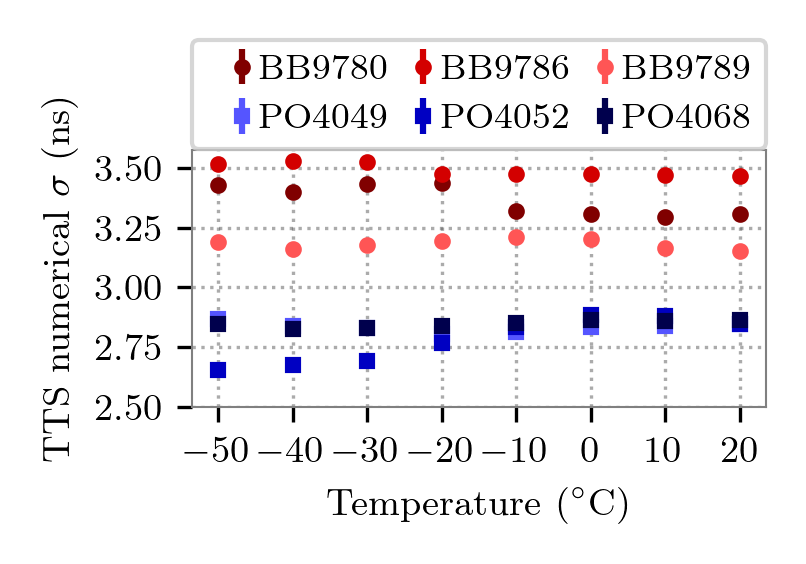}
     \end{subfigure}
        \caption{Evaluation for the TTS: (left) Binned arrival times, with the orange line indicating a Gaussian fit around the peak of the distribution and the blue line representing the data. (right) Comparison of the numerically calculated TTS plotted against temperature for each PMT.}
        \label{fig:TTS}
\end{figure}

For the TTS analysis, the relative pulse times with respect to the light source trigger are plotted in a histogram as shown in \cref{fig:TTS} (left). The distribution of the histogram deviates from a Gaussian shape, particularly in the right \textit{tail} region which is a consequence of full photocathode illumination in conjunction with the asymmetric curvature of the first dynode. While this is observed for all PMTs it is more pronounced in the models studied in this work. Hence, instead of fitting a Gaussian distribution to the peak, the numerical standard deviation of the entire main peak distribution (-8\,ns, 17\,ns) is used (\cref{fig:TTS} (right)). %The results for the TTS are presented in \cref{fig:TTS} (right). 
Overall, NNVT PMTs feature on average about ($18.6 \pm 0.1$)\,\% better time resolution compared to Hamamatsu PMTs.
%Overall, NNVT PMTs feature a better time resolution than Hamamatsu PMTs with an average TTS decrease of ($18.62 \pm 0.09$)\,\%. 
PMT BB9780 shows an increase of 3.5\,\% in TTS towards lower temperatures whilst PO4052 demonstrates an 8\,\% decrease. The other PMTs perform essentially constant (<1.5\,\% relative deviations).  %NNVT 0.0602 HPK 0.1327
Though statistics is low Hamamatsu PMTs tend to have stronger variations in TTS, with comparatively twice the overall standard deviation.

\section{Quantum efficiency}\label{sec:qe}

\begin{wrapfigure}{r}{0.4\textwidth}
\vspace{-20pt}
  \begin{center}
    \includegraphics[width=0.4\textwidth]{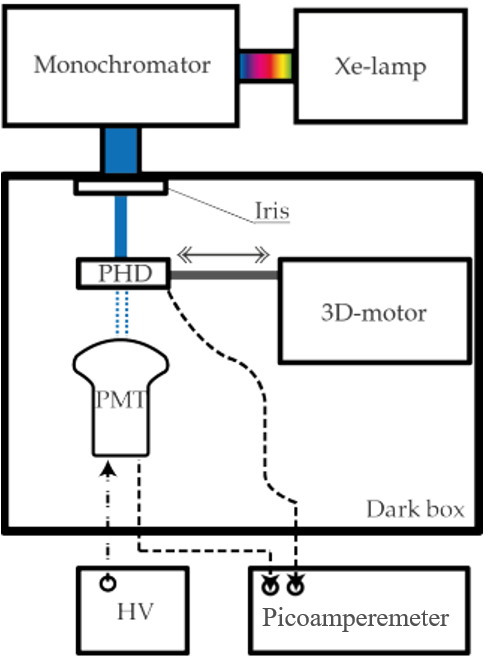}
  \end{center}
  \vspace{-20pt}
  \caption{Experimental setup to measure the quantum efficiency. Taken from \cite{MartinDoctorialThesis}.}
  \label{fig:QEsetup}
  %\vspace{-20pt}
\end{wrapfigure}

\noindent
The quantum efficiency (QE) is a crucial PMT characteristic. It represents the probability that a photoelectron is emitted by a photon striking the photocathode, and is wavelength dependent. For this, the setup shown in \cref{fig:QEsetup} is used.
The wavelength $\lambda$ of high intensity light (emitted from a Xenon lamp) is selected by a monochromator in the range of 250\,nm -- 700\,nm in increments of 10\,nm. %which is guided into a light-tight enclosure. 
The beam is directed either to a calibrated photodiode (PHD in \cref{fig:QEsetup}) or a PMT. The PMT is attached to a base that shortcuts all dynodes, allowing collection of each photoelectron striking the multiplier system without multiplication. 
%Since the beam is divergent, either the entire photocathode ($\sim$ 80\,mm) or a small spotlight ($\sim$ 3\,mm) can be illuminated depending on the distance. 
Since the beam is divergent, its spot size on the PMT can be adjusted to illuminate the entire photocathode by positioning the PMT further from the iris.
%For more information, the reader is directed to \cite[p.~30]{MarkusMasterThesis}.
The respective current $I$ as well as the dark current $DC$ are measured using a picoamperemeter. The quantum efficiency of the PMT is calculated using the equation
\begin{equation*}
    QE(\lambda) = \frac{I_{\text{PMT}}(\lambda)-DC_{\text{PMT}}}{I_{\text{Diode}}(\lambda)-DC_{\text{Diode}}} \cdot QE_{\text{Diode}}(\lambda).
\end{equation*}

\begin{figure}[!htb]
     \centering
     \begin{subfigure}[!hb]{0.45\textwidth}
         \centering
         \includegraphics[width=\textwidth]{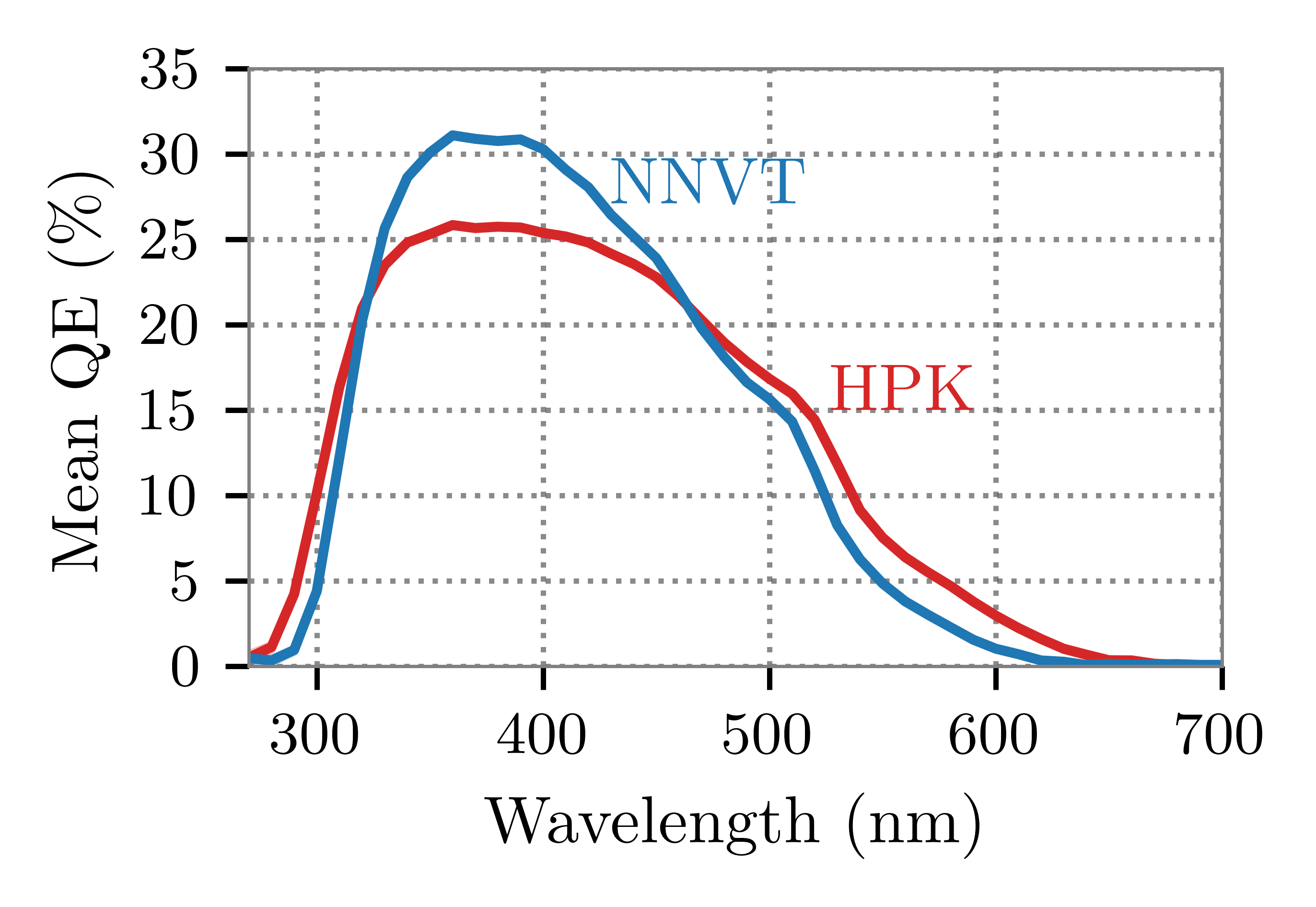}
     \end{subfigure}
     \hfill
     \begin{subfigure}[!hb]{0.45\textwidth}
         \centering
         \includegraphics[width=\textwidth]{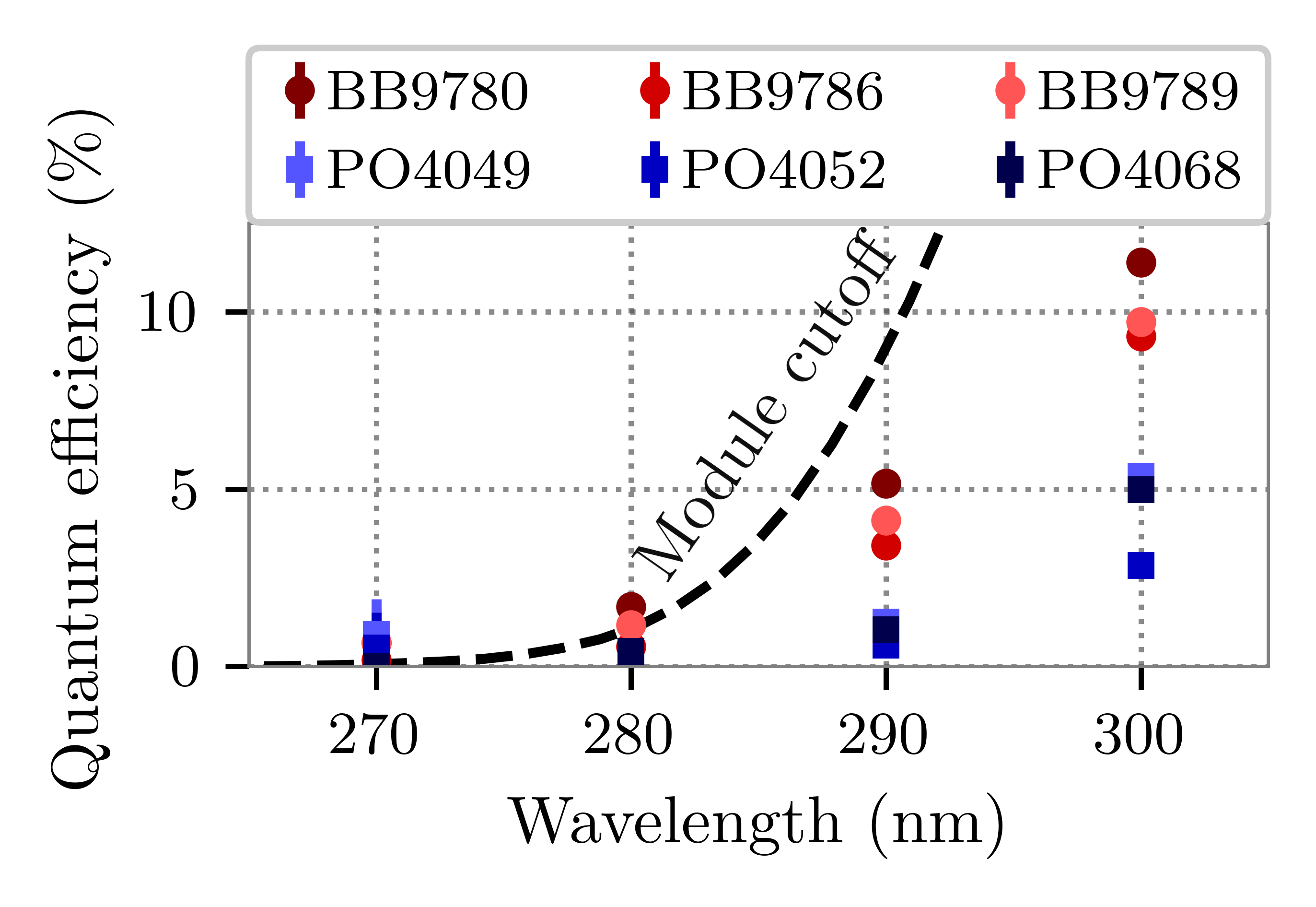}
     \end{subfigure}
        \caption{Quantum efficiency for the entire spectral range (left) and near the module cutoff (right).}
        \label{fig:QE}
\end{figure}

\noindent
The results are shown in \cref{fig:QE}. 
%NNVT PMTs feature a higher maximum quantum efficiency at wavelengths around 360\,nm (($31.1 \pm 0.1$)\% vs. ($25.8 \pm 0.1$)\%) whereas Hamamatsu PMTs exhibit a slightly broader spectrum. %and can feature \textit{high QE} as well for extra costs. 
NNVT PMTs feature a higher maximum quantum efficiency at wavelengths around 360\,nm (($31.1 \pm 0.1$)\%) than Hamamatsu PMTs (($25.8 \pm 0.1$)\%) which exhibit a slightly broader spectrum.
Both PMT models feature a similar cutoff at around 270\,nm since both PMT corpora consist of borosilicate glass. The same holds true for the pressure vessel of IceCube-Gen2s optical module as indicated by the black line in the right plot where a comparable PMT quantum efficiency is desired. Hamamatsu PMTs tend to show slightly higher quantum efficiencies near this cutoff which may be attributed to thinner glass in front of the photocathode area.

\section{Homogeneity scan}\label{sec:scan} %Performance with respect to photocathode position
\noindent
Depending on the incident position of the photon on the photocathode, the aforementioned characteristics may change due to different path lengths of photoelectrons inside of a PMT.% due to the complex shaped electric potential. 

\begin{wrapfigure}{r}{0.45\textwidth}
\vspace{-20pt}
  \begin{center}
    \includegraphics[width=0.45\textwidth]{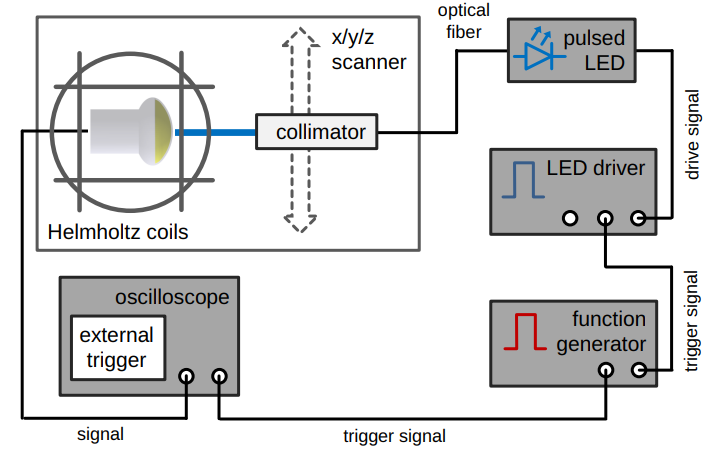}
  \end{center}
  \vspace{-20pt}
  \caption{Experimental setup to scan the photocathode. Taken from \cite{MartinScan}.}
  \label{fig:Scansetup}
  \vspace{-20pt}
\end{wrapfigure}

Instead of using diffuse illumination, light from a collimated fiber is used to scan the photocathode surface (\cref{fig:Scansetup}). %This provides valuable information on the spatial dependence of these parameters and helps to assess the homogeneity of the PMT’s performance.
The PMT is located inside a Helmholtz-cube which is set to compensate Earth's magnetic field, nullifying its influence on the measurements, whereas the data acquisition is analogous to \cref{sec:temp}. For additional information, the reader is referred to \cite{MartinScan}.

\begin{figure}[!htb]
     \centering
     \begin{subfigure}[!htb]{0.45\textwidth}
         \centering
         \includegraphics[width=\textwidth]{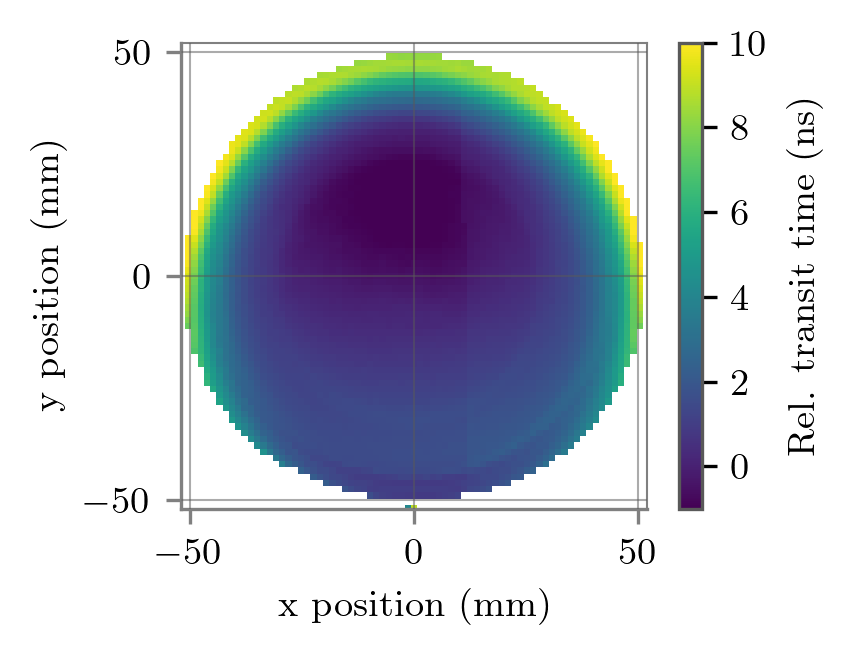}
     \end{subfigure}
     \hfill
     \begin{subfigure}[!htb]{0.45\textwidth}
         \centering
         \includegraphics[width=\textwidth]{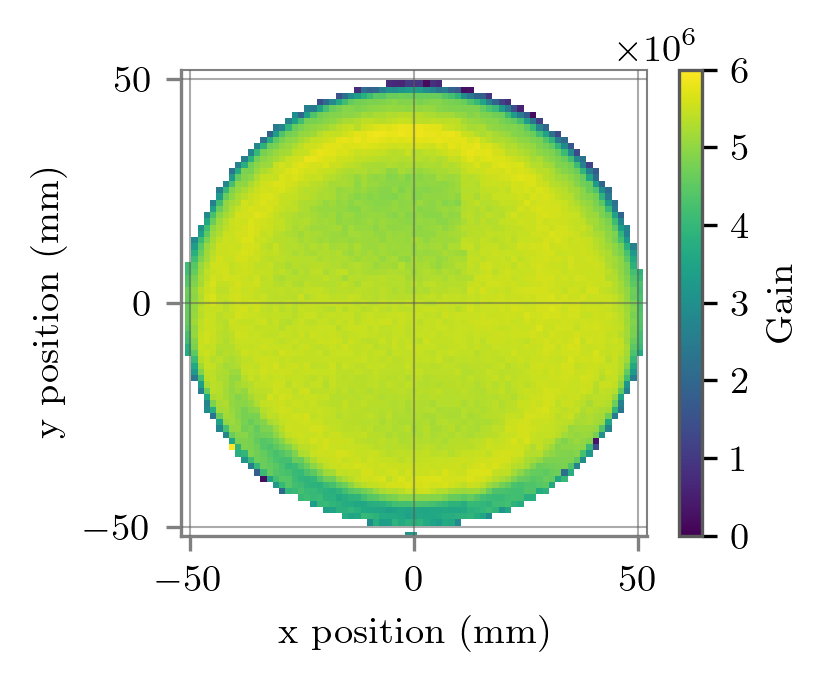}
     \end{subfigure}
        \caption{Entire data set for relative transit time (left) and absolute gain (right) for BB9786.}
        \label{fig:ScanMaps}
\end{figure}

In the following, the center region is defined for the area with $r<30$\,mm and the edge region for the area with $r>45$\,mm.
The resulting scan of the transit time relative to the center region and the absolute gain is shown in \cref{fig:ScanMaps}. It is apparent that the performance deteriorates towards the edges which is averaged over in the diffuse light measurements as presented in \cref{sec:temp}. The asymmetric distribution of the transit time values arises from the first dynode being curved in the positive y-direction in this representation.

\begin{figure}[!htb]
     \centering
     \begin{subfigure}[!htb]{0.45\textwidth}
         \centering
         \includegraphics[width=\textwidth]{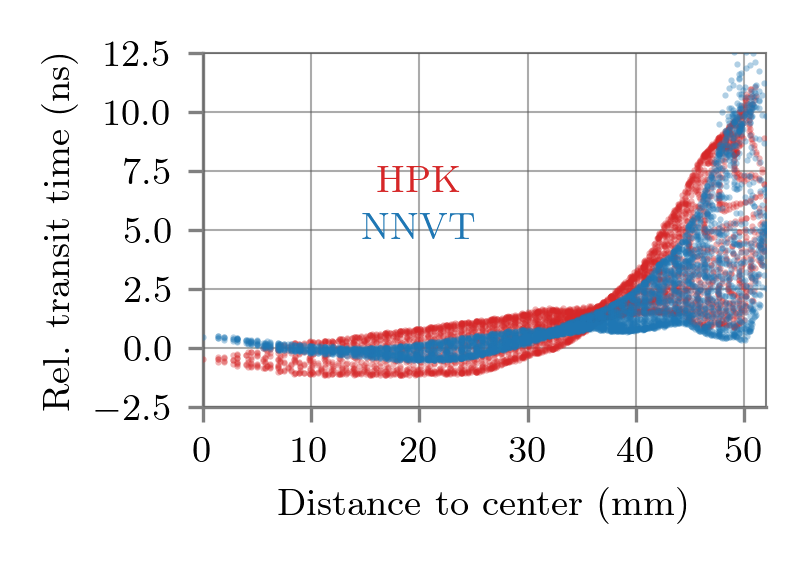}
     \end{subfigure}
     \hfill
     \begin{subfigure}[!htb]{0.45\textwidth}
         \centering
         \includegraphics[width=\textwidth]{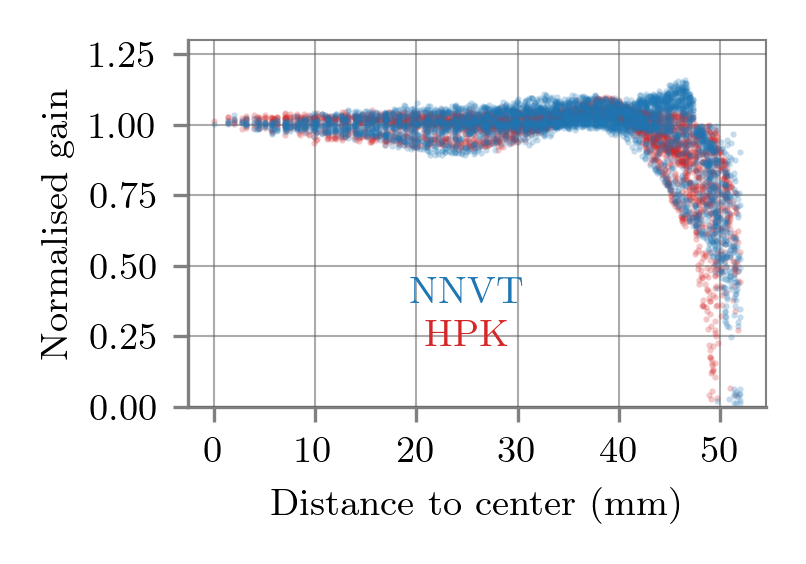}
     \end{subfigure}
        \caption{Comparison between both PMT models as a function of radial distance towards center. \linebreak (left) Transit time relative to the center. (right) Gain normalised to the center.}
        \label{fig:ScanCompare}
\end{figure}

While the gain values demonstrate relatively homogeneous behavior across the radial distance,  the transit time values exhibit a noticeable slope along the y-direction.
%Glass smudges on NNVT...surface not as smooth, more reflections...
In order to compare both PMT models, instead of heatmaps as presented in \cref{fig:ScanMaps}, the results in \cref{fig:ScanCompare} are shown as scatter plots as a function of radial distance to the photocathode center.

\noindent Comparing both data sets, the mean and standard deviations for the overall transit time are \linebreak ($1.75 \pm 2.48$)\,ns for NNVT and ($2.10 \pm 2.66$)\,ns for Hamamatsu PMTs whereas the edge regions exhibit values of ($5.37 \pm 3.46$)\,ns and ($5.93 \pm 3.07$)\,ns respectively, while NNVT PMTs show slightly higher maximum deviations but a more uniform distribution in the center. %Even though Hamamatsu PMTs are operated at a higher nominal voltage, 
Thus, NNVT PMTs demonstrate a more uniform timing response in the center but worse towards the edges. 
Contrarily, regarding the gain, the standard deviations relative to the mean in the center are 3.7\,\% (NNVT) and 3.1\,\% (Hamamatsu), whereas at the edges, they are 20.5\,\% (NNVT) and 19.0\,\% (Hamamatsu). %Consequently, Hamamatsu PMTs exhibit approximately 6\,\% better homogeneity in the center region and an 18\,\% lower deviation near the edges. 
Smoothed curves reveal a drop of 20\% in gain at approximately $0.93 \cdot r$ for both models and a slight increase in gain is observed around 40\,mm. %which may stem from internal reflections.
In both regards the investigated 4-inch PMTs exhibit similar performance and are within the expectations. However, it is important to note that only a single PMT from each vendor was measured.

\section{Conclusion}\label{sec:final}
\noindent
Considering the importance of the PMTs' outer dimensions for module construction (refer to \cite{Yuya:2023icrc}), the performance requirements have a certain degree of flexibility.
The metrics presented in this work are listed in \cref{tab:final}, and both manufacturers fulfill the specifications set by the collaboration. In summary, NNVT PMTs operate at lower nominal voltages and exhibit higher dark rates due to correlated noise, with larger PMT-to-PMT fluctuations compared to Hamamatsu PMTs. NNVT PMTs offer better timing resolution and higher quantum efficiency, though with a slightly higher cutoff. Whilst NNVT PMTs are slightly more homogeneous in the PMT center, the PMT response exhibits more variation towards the edges. Although this study had a limited sample size and more PMTs need to be measured, the results suggest that the tested 4-inch NNVT PMTs performed slightly better compared to Hamamatsu PMTs.
To conclude, both vendors are viable options for the optical modules of IceCube-Gen2, which require production of over 160,000 PMTs. %Hamamatsu dark rates high with mass production melting furnace. Both struggle with the enorm order volume.

%I had to kick out SPE resolution, Peak-To-Valley ratio, fall time, rise time, pulse type probabilities (all measure temperature dependant) since it would be too crowded
\begin{table}[!htb]
\centering
\adjustbox{max width=\textwidth}{%
\begin{tabular}{l|ccccccc}
& Req.  & BB9780 & BB9786 & BB9789 & PO4049 & PO4052 & PO4052 \\
\hline
Nominal  Voltage (V)        & <\,1500    &  994  &  996  &  995  &  741  &  728  &  828  \\
%TTS (ns)       &  3.4  &  3.37  &  3.49  &  3.18  &  2.84  &  2.78 & 2.85 \\ %numerical ... requirement stated in FWHM
TTS (ns)       &  <\,3.4  &  2.51  &  2.58  &  2.38  &  1.75  &  1.86 & 1.75 \\ %From gaussian peak fit
%Rise Time ns)        &  <\,5  &    &    &    &    &    \\
%Peak to Valley ratio    &  >\,2   &    &    &    &    &    \\
Max QE (\%)       &  >\,25  &  25.95  &   26.24 &  25.29  &  29.85  &  30.15 &  33.24 \\
%Pre/late/after (\%) & <\,1/5/15 &    &    &    &    &    &   
\end{tabular}}
\caption{PMT requirements considered in this work. The TTS is defined as 8\,ns (FWHM), and the values stated here represent the standard deviations of a Gaussian fit to the peak of the main peak timing histogram.}
\label{tab:final}
\end{table}

% Bibtex references:
\bibliographystyle{ICRC}
\bibliography{references}

\clearpage

%The following list of authors, affiliations and funding agencies will be updated at the day of submission. The following template is a placeholder generated via https://authorlist.icecube.wisc.edu/icecube on March 25, 2023 and will be updated.
%\input{authorlist_IceCube.tex}

\section*{Full Author List: IceCube-Gen2 Collaboration}

\scriptsize
\noindent
R. Abbasi$^{17}$,
M. Ackermann$^{76}$,
J. Adams$^{22}$,
S. K. Agarwalla$^{47,\: 77}$,
J. A. Aguilar$^{12}$,
M. Ahlers$^{26}$,
J.M. Alameddine$^{27}$,
N. M. Amin$^{53}$,
K. Andeen$^{50}$,
G. Anton$^{30}$
% ...
\\
\\
% ...
$^{12}$ Universit{\'e} Libre de Bruxelles, Science Faculty CP230, B-1050 Brussels, Belgium \\
% ...
$^{17}$ Department of Physics, Loyola University Chicago, Chicago, IL 60660, USA \\
% ...
$^{22}$ Dept. of Physics and Astronomy, University of Canterbury, Private Bag 4800, Christchurch, New Zealand \\
% ...
$^{26}$ Niels Bohr Institute, University of Copenhagen, DK-2100 Copenhagen, Denmark \\
$^{27}$ Dept. of Physics, TU Dortmund University, D-44221 Dortmund, Germany \\
% ...
$^{30}$ Erlangen Centre for Astroparticle Physics, Friedrich-Alexander-Universit{\"a}t Erlangen-N{\"u}rnberg, D-91058 Erlangen, Germany \\
% ...
$^{47}$ Dept. of Physics and Wisconsin IceCube Particle Astrophysics Center, University of Wisconsin{\textendash}Madison, Madison, WI 53706, USA \\
% ...
$^{50}$ Department of Physics, Marquette University, Milwaukee, WI, 53201, USA \\
% ...
$^{53}$ Bartol Research Institute and Dept. of Physics and Astronomy, University of Delaware, Newark, DE 19716, USA \\
% ...
$^{76}$ Deutsches Elektronen-Synchrotron DESY, Platanenallee 6, 15738 Zeuthen, Germany  \\
$^{77}$ Institute of Physics, Sachivalaya Marg, Sainik School Post, Bhubaneswar 751005, India \\
% ...

\subsection*{Acknowledgements}

\noindent
The authors gratefully acknowledge the support from the following agencies and institutions:
USA {\textendash} U.S. National Science Foundation-Office of Polar Programs,
U.S. National Science Foundation-Physics Division,
U.S. National Science Foundation-EPSCoR,
Wisconsin Alumni Research Foundation,
Center for High Throughput Computing (CHTC) at the University of Wisconsin{\textendash}Madison,
Open Science Grid (OSG),
Advanced Cyberinfrastructure Coordination Ecosystem: Services {\&} Support (ACCESS),
Frontera computing project at the Texas Advanced Computing Center,
U.S. Department of Energy-National Energy Research Scientific Computing Center,
Particle astrophysics research computing center at the University of Maryland,
Institute for Cyber-Enabled Research at Michigan State University,
and Astroparticle physics computational facility at Marquette University;
Belgium {\textendash} Funds for Scientific Research (FRS-FNRS and FWO),
FWO Odysseus and Big Science programmes,
and Belgian Federal Science Policy Office (Belspo);
Germany {\textendash} Bundesministerium f{\"u}r Bildung und Forschung (BMBF),
Deutsche Forschungsgemeinschaft (DFG),
Helmholtz Alliance for Astroparticle Physics (HAP),
Initiative and Networking Fund of the Helmholtz Association,
Deutsches Elektronen Synchrotron (DESY),
and High Performance Computing cluster of the RWTH Aachen;
Sweden {\textendash} Swedish Research Council,
Swedish Polar Research Secretariat,
Swedish National Infrastructure for Computing (SNIC),
and Knut and Alice Wallenberg Foundation;
European Union {\textendash} EGI Advanced Computing for research;
Australia {\textendash} Australian Research Council;
Canada {\textendash} Natural Sciences and Engineering Research Council of Canada,
Calcul Qu{\'e}bec, Compute Ontario, Canada Foundation for Innovation, WestGrid, and Compute Canada;
Denmark {\textendash} Villum Fonden, Carlsberg Foundation, and European Commission;
New Zealand {\textendash} Marsden Fund;
Japan {\textendash} Japan Society for Promotion of Science (JSPS)
and Institute for Global Prominent Research (IGPR) of Chiba University;
Korea {\textendash} National Research Foundation of Korea (NRF);
Switzerland {\textendash} Swiss National Science Foundation (SNSF);
United Kingdom {\textendash} Department of Physics, University of Oxford.\\

\noindent
This work was supported by the German Bundesministerium für Bildung und Forschung (BMBF) Verbundforschung grants 05A20PM2.

\end{document}